\documentclass[twocolumn,secnumarabic,amssymb,graphics,floatfix,nofootinbib,tightenlines,nobibnotes,aps,notitlepage,superscriptaddress,10pt]{revtex4-1}

\usepackage{graphicx}% Include figure files
\usepackage{dcolumn}% Align table columns on decimal point
\usepackage{bm}% bold math
\usepackage{bigints}
\usepackage{hyperref}
\hypersetup{colorlinks=true}
\usepackage{xcolor}  % for writting parts in color, e.g. in red

\usepackage{subdepth}

\begin{document}

\title{Critical phenomena and nonlinear dynamics in a spin ensemble \\ strongly coupled to a cavity. I. Semiclassical approach}

\author{Dmitry O. Krimer}
\email[]{dmitry.krimer@gmail.com}
\author{Matthias Zens}
\author{Stefan Rotter}
\affiliation{Institute for Theoretical Physics, Vienna University of Technology (TU Wien), Wiedner Hauptstra\ss e 8-10/136, A--1040 Vienna, Austria, EU}

\begin{abstract}
We present a theoretical study on the nonlinear dynamics and stationary states of an inhomogeneously broadened spin ensemble coupled to a single-mode cavity driven by an external drive with constant amplitude. Assuming a sizeable number of constituents within the ensemble allows us to use a semiclassical approach and to formally reduce the theoretical description to the Maxwell-Bloch equations for the cavity and spin amplitudes. We explore the critical slowing-down effect, quench dynamics, and asymptotic behavior of the system near a steady-state dissipative phase transition accompanied by a bistability effect. Some of our theoretical findings have recently been successfully verified in a specific experimental realization based on a spin ensemble of negatively charged nitrogen-vacancy centers in diamond strongly coupled to a single-mode microwave cavity (see \href{https://doi.org/10.1126/sciadv.1701626}{Science Adv. {\bf 3}, e1701626 (2017)}).
\end{abstract}

\maketitle

\section{Introduction}

The phenomenon of optical bistability has been extensively analyzed and experimentally realized in various systems exhibiting driven-dissipative phase transitions such as a laser with a saturable absorber \cite{Lugiato:1978ab},  resonant cavities of different shapes  filled with two-level atoms  \cite{Lugiato:1978aa, Abraham:1980ab,Abraham:1980aa,Hassan:1978aa} or an interferometer with a nonlinear absorber \cite{Drummond:1981aa} driven by the coherent incident field -- to name just a few examples (see also \cite{Lugiato:1984ab} for a review). In all of these systems, the incident field was either on resonance or close to resonance with the cavity field -- the two different cases referred to, respectively, as absorptive or dispersive optical bistability. At the same time, it was shown that while the bistability effect with hysteresis arises on the semiclassical level when the problem is formally reduced to the Maxwell-Bloch equations, a quantum treatment of the coherently driven cavity predicts a unique quantum steady solution for the cavity amplitude without amplitude bistability \cite{Drummond:1980aa}. The uniqueness of the quantum state was attributed to quantum or classical fluctuations which lead to a finite probability for a system to jump from one stable branch to another and, as a consequence, to smearing out the bistability region \cite{Risken:1987aa}. However, in the thermodynamic limit, these fluctuations are negligibly small, so that the semiclassical solution featuring the bistability effect is well-justified for many experimental realizations \cite{Lugiato:1984ab}. Two natural questions which arose in this context are the following: how many atoms for the onset of the thermodynamic limit are required and what kind of system characteristics can trigger this onset \cite{Rempe:1991aa}?

Over the last decade, open quantum systems featuring a driven-dissipative phase transition in the thermodynamic limit became a subject of renewed interest caused by technological progress in various setups of cavity quantum electrodynamics (QED). Many of these setups are studied in terms of their potential for the storage and processing of quantum information, secure communication and quantum sensing. Particularly attractive in this context are so-called ``hybrid quantum systems'' (HQS), which conflate the individual advantages of different quantum technologies \cite{Kurizki:2015aa}. Among recent realizations of such HQS, those based on spin, atomic, or molecular ensembles coupled to superconducting microwave cavities have attracted broad attention \cite{Xiang:2013aa}. The technology for building such devices has meanwhile advanced up to the degree that state-of-the-art experiments can be performed on a single superconducting chip, on which the corresponding ensemble is probed through the in- and out-coupling of a microwave field. Other interesting alternatives for HQS are being realized with opto- and nano- mechanical systems, which enable the conversion between microwave and optical photons via phonons \cite{Aspelmeyer:2014aa}.

The experimental success achieved in various physical realizations led to intensive theoretical studies in this rapidly developing field. Many intriguing effects on both sides of the of the semiclassical-to-quantum boundary have recently been observed in a relatively simple system comprising a single-mode driven dissipative cavity with a Kerr optical nonlinearity.  Among them are, e.g., dynamic hysteresis effects \cite{Casteels:2016aa, Rodriguez:2017aa}, nonadiabatic effects under periodic driving \cite{Reimer:2018aa}, photon blockade and multiphoton resonant effects under additional parametric driving \cite{Bartolo:2016aa}, and the emergence of a jump in the observable when proceeding to the thermodynamic limit \cite{Minganti:2018aa}. Recent theoretical studies also highlight the importance of metastable states in the understanding of a driven dissipative phase transition \cite{Minganti:2018aa,Plenio:2017aa,Macieszczak:2016aa} and the geometrical nature of the metastable dynamics \cite{krimer:2018ac}. Strong coupling of a single spin or spin ensemble to the cavity leads to the emergence of a rich variety of other intriguing phenomena like the breakdown of the photon blockade for increasing drive power \cite{Carmichael:2015aa,Fink:2017aa}, the bistability effect for just a few atoms \cite{Dombi:2013aa} or for extremely low saturation photon numbers \cite{Dombi:2015aa} as well as bistable versus metastable behavior in driven dissipative Rydberg gases \cite{Martin:2011aa}.

In a recent paper \cite{Angerer:2017aa} involving also the present authors among others, it was shown how a hybrid system composed of a superconducting resonator coupled to an inhomogeneously broadened spin ensemble in diamond can be used to explore amplitude bistability in new regimes of cavity QED. In accordance with first theoretical findings based on a semiclassical approach, the experiment demonstrated a critical slowing down of the cavity population on the order of eleven hours - a timescale much longer than observed anywhere else for this phenomenon. In this paper we present a detailed theoretical study of this effect including an analysis based on adiabatically eliminating the cavity field and a description of the asymptotic behavior for long times as well as of absorptive optical bistability. The system under study consists of an inhomogeneously broadened spin ensemble strongly coupled to a single-mode resonator exhibiting a driven dissipative phase transition. We consider the thermodynamic limit justified for a very large number of spins where the problem can be treated in the framework of the Maxwell-Bloch equations. In accordance with the recent experimental realization \cite{Angerer:2017aa}, the cavity decay rate and the non-radiative dephasing of individual spins are chosen to be much larger than their radiative dephasing. Among other phenomena mentioned above, we find an interesting separation of time scales in the temporal dynamics. Specifically, the system exhibits damped Rabi oscillations of the cavity amplitude and small spin deviations from the spins' unexcited ground state at short times, even for moderate values of the driving amplitude. The transient state that the system sets into is reminiscent of a stationary state for a relatively long time span. However, at even longer timescales, the system deviates from this transient state towards its ultimate stationary state with possible large spin deviations from the ground state, characterized by a strong dependence on the driving amplitude.

All issues related to the onset of the thermodynamic limit are addressed in a separate companion paper, where we use a cumulant-expansion approach to analyze the semiclassical-to-quantum boundary as the size of the spin ensemble grows \cite{Zens:2019aa}. 

Our paper is organized as follows. In Sec.~\ref{Sec_Theory}, we present the theoretical framework of our problem summarizing all important assumptions made in our approach. In Sec.~\ref{Dynamics_bistab}, we consider the dynamics and bistability effects under the action of an external drive with constant amplitude for different shapes of the spectral spin distribution. Section~\ref{Sec_Adiab_elim} is devoted to adiabatic elimination of the cavity amplitude and critical slowing down phenomena. In Sec.~\ref{Sec_asympt_decay}, we investigate asymptotic decay towards a stationary state. Finally, we draw our conclusions in Sec.~\ref{Sec_Concl}.

\section{Theoretical model}
\label{Sec_Theory}

Our starting point is the Tavis-Cummings Hamiltonian ($\hbar=1$) \cite{Tavis:1968aa}
\begin{eqnarray}
&&{\cal H}=\Delta_c a^{\dagger}a+\frac{1}{2}\sum_k^N\Delta_k\sigma_k^z+\sum_k^N\left[g_k\sigma_k^-a^{\dagger}+g_k^*\sigma_k^+a\right]
\nonumber\\
&&+i\left[\eta(t) a^{\dagger}-\eta^*(t) a\right]\,,
\label{Hamilt_fun}
\end{eqnarray}
where $a^{\dag}$ and $a$ are standard creation and annihilation operators of the single cavity mode with frequency $\omega_c$ and $\sigma_k^+,\,\sigma_k^-,\,\sigma_k^z$ are the Pauli operators associated with each individual spin of frequency $\omega_k$. Here $\Delta_k=\omega_k-\omega_p$ and $\Delta_c=\omega_c-\omega_p$ are detunings with respect to the external driving frequency $\omega_p$ and $g_k$ stands for the coupling strength of the $k$-th spin. An incoming signal is characterized by the carrier frequency $\omega_p$ and by the amplitude $\eta(t)$. The interaction part of ${\cal H}$ is written in the dipole and rotating-wave approximation (terms $\propto a\sigma_k^-,\,a^\dag \sigma_k^+$ are neglected).

A quantum master equation \cite{Breuer:2007aa} for the spin-cavity density matrix can be written in the following form, $d\rho/dt=-i\left[\,{\cal H},\rho\,\right]\,+\,\mathcal{L}_D(\rho)$, where ${\cal H}$ stands for the Hamiltonian (\ref{Hamilt_fun}) and $\mathcal{L}_D(\rho)$ is the standard Lindblad operator which accounts for the system-environment interaction as follows,
\begin{eqnarray}
\nonumber
&&\!\mathcal{L}_D(\rho)\!=\!\kappa\,(2a\rho a^\dagger-a^\dagger a\,\rho-\rho\, a^\dagger a)+\gamma_p\sum\limits_{j=1}^N(\sigma_j^z\rho\,\sigma_j^z-\rho\,)
\\[6pt]
&&+\gamma_h\sum\limits_{j=1}^N(2\sigma_j^-\rho\,\sigma_j^+-\sigma_j^+\sigma_j^-\rho-\rho\,\sigma_j^+\sigma_j^-),
\label{eq_Lindblad}
\end{eqnarray}
where the first term describes the cavity losses with the decay rate $\kappa$. We will explicitly distinguish between non-radiative dephasing (second term) and radiative dephasing (third term) of the individual spins characterized by two different relaxation rates, $\gamma_p$ and $\gamma_h$, respectively. Using this formalism, one can derive a first-order linear ordinary differential equation (ODE) for the expectation value of any operator $O$, which is given by, $d\langle O\rangle/dt=\, \text{Tr}\!\left\{-i[\,O,{\cal H}\,]\rho\,+ O\mathcal{L}_D(\rho)\right\}$. In what follows, we study the thermodynamic (semiclassical) limit where the number of spins is taken to infinity ($N\rightarrow \infty$) and all correlations between spin and cavity operators are neglected, i.e., the second-order expectation values like $\langle \sigma_k^- a^\dag \rangle$ and $\langle \sigma_k^z a \rangle$ factorize into products of the first-order expectation values $\langle \sigma_k^-\rangle \langle a^\dag \rangle$ and $\langle \sigma_k^z \rangle\langle a \rangle$. With these approximations we arrive at the well-known Maxwell-Bloch equations  \cite{Bonifacio:1982aa} for the cavity and spin expectation values, $\langle a \rangle$,  $\langle \sigma_k^-\rangle$ and $\langle\sigma_k^z\rangle$, which form a closed set of nonlinear equations
\begin{subequations}
\begin{eqnarray}
\label{Eq_a_Volt}
&&\dot{a}= -(\kappa+i\Delta_c)a -i \sum\nolimits_k g_k  \sigma_k^-+\eta(t), 
\\[6pt]
\label{Eq_bk_Volt}
&&\dot{\sigma}_k^- = -(\gamma_\perp+i \Delta_k)\sigma_k^- +ig_k  \sigma_k^z a,
\\[6pt]
\label{Eq_ck_Volt}
&&\dot{\sigma}_k^z = -\gamma_{\parallel}(1+ \sigma_k^z)+2i g_k ( \sigma_k^-a^{\dag}- \sigma_k^+ a),
\end{eqnarray}
\end{subequations}
where $\gamma_\perp=\gamma_h+2\gamma_p$ and $\gamma_{\parallel}=2\gamma_h$  are the transverse and longitudinal relaxation rates, respectively. For simplicity we omit in Eqs.~(\ref{Eq_a_Volt}-\ref{Eq_ck_Volt}) and everywhere below the angle brackets $\langle \ldots \rangle$ that indicate expectation values. The nonlinear nature of  Eqs.~(\ref{Eq_a_Volt}-\ref{Eq_ck_Volt}) notably stems from the above-mentioned factorization procedure applied to intrinsically linear ODEs for the expectation values.

In what follows the main focus of our study will be the dynamics in systems for which non-radiative processes constitute the dominant dephasing mechanism, i.e., $\gamma_h \ll \gamma_p$ and therefore: $\gamma_\parallel \ll \gamma_{\perp}$. Moreover, we also assume that the cavity decay rate $\kappa$ is orders of magnitude larger than the longitudinal relaxation rate, so that the following inequality holds in addition: $\gamma_\parallel \ll \kappa$. These two inequalities are very well fulfilled as, e.g., in the recent experiments mentioned in the introduction \cite{Putz:2014aa,Angerer:2017aa}.

For many physical realizations the individual spin coupling strengths $g_k$ or/and spin frequencies $\omega_k$ are not the same but rather distributed around certain mean values - an effect commonly referred to as inhomogeneous broadening of the spin ensemble. As mentioned above, the thermodynamic limit is justified for a sizeable number of constituents (spins, qubits etc.) in which case the distribution of coupling strengths and/or of spin frequencies is a smooth function around the mean value. Such a phenomenological continuous spectral spin density allows one to conveniently treat the problem in the framework of a Volterra equation for the cavity amplitude $a$ \cite{Putz:2014aa, Krimer:2014aa}, valid in the limit of weak driving signals $\eta$, when the so-called Holstein-Primakoff-approximation holds \cite{Primakoff:1939aa}. The specific shape of the spin density depends on the physical system under study and can typically be determined by a careful comparison with the experiment based on stationary or dynamical transmission measurements. 

Following previous studies \cite{Sandner:2012aa, Putz:2014aa, Krimer:2014aa}, we model the shape of the spin spectral density, $\rho(\omega)=\sum_k g_k^2 \delta(\omega-\omega_k)/\Omega^2\!$, by a $q$-Gaussian distribution which is symmetric with respect to the mean frequency $\omega_s$,
\begin{eqnarray}
\label{rho_w_Eq}
\rho(\omega)=B\left[1-(1-q)(\omega-\omega_s)^2/\Delta^2\right]^{1/(1-q)},
\end{eqnarray}
where $q$ is the dimensionless shape parameter, $1<q<3$, $\gamma_q=2\Delta\sqrt{(2^q-2)/(2q-2)}$ is the full-width at half maximum (FWHM), and $B$ is a normalization constant. Note that Gaussian and Lorentzian distributions are recovered for $q\rightarrow 1$ and  $q=2$, respectively. The parameter $\Omega$ introduced right before Eq.~(\ref{rho_w_Eq}) in the formal expression for the spectral spin density is the collective coupling strength of the spin ensemble to the cavity, $\Omega^2=\sum_j^Ng_j^2$. It is worth noting that $\Omega$ scales with the ensemble size as $\sqrt{N}$ so that for large spin ensembles its value can easily exceed the total decoherence rate of the system giving rise to the strong-coupling regime (see, e.g., \cite{Amsuss:2011aa,Kubo:2011aa} for nitrogen-vacancy (NV) spin ensembles).
 
Importantly, in the general case when the initially unexcited spins are driven away from the south pole of the Bloch sphere through a sizable driving amplitude $\eta$, the dynamics becomes essentially nonlinear and it is no longer possible to formulate the problem in the simple form of a Volterra equation. Neither will it be possible to solve the problem in the continuous limit. Instead, similar to previous work \cite{Angerer:2017aa}, we discretize the spectral density by performing the inverse transformation, $g_j=\Omega \left[\rho(\omega_j)/\sum_{l=1}^M\rho(\omega_l)\right]^{1/2}$, where the shape of $\rho(\omega)$ can be determined at the stage of linear dynamics governed by the Volterra equation. In other words, we make our problem numerically tractable by dividing the entire frequency interval into $M$ clusters of equal size, where each cluster is characterized by the coupling strength $g_j$. Thus, each $g_j$ effectively represents the coupling strength of a ``large'' spin residing in the $j$-th cluster within the frequency subinterval $\omega_j$ to $\omega_j+\Delta\omega_j$ rather than an individual coupling strength. Another possible way of mapping the continuous to the discrete case is to keep both the (frequency) size of each cluster and the coupling strength to each spin the same, but filling up each cluster with a different number of spins distributed in accordance with the shape of $\rho(\omega)$.

If not specified otherwise, our numerical calculations are performed with a set of parameters typical for the experiments with a $\lambda/2$ superconducting microwave coplanar waveguide resonator magnetically coupled to a spin ensemble of negatively charged NV centers in diamond \cite{Angerer:2017aa,Putz:2014aa,Krimer:2014aa}: the cavity decay rate $\kappa/2\pi=0.8$\,MHz, the coupling strength $\Omega/2\pi=12$~MHz,  the transverse spin relaxation rate $\gamma_{\perp}/2\pi=250$~kHz, the dimensionless $q$-Gaussian parameter $q=1.39$, and the FWHM of the $q$-Gaussian $\gamma_q/2\pi=9.4$~MHz. The mean frequency of the spectral density, the cavity frequency and the probe frequency of the driving signal are all taken to be in resonance, $\omega_s=\omega_c=\omega_p=2\pi\times 2.6915$\,GHz. Next, the value for the longitudinal relaxation rate, $\gamma_{\parallel}/2\pi=100$~Hz, is chosen to be much smaller than the transverse one, $\gamma_{\perp}$, but still at least two orders of magnitude larger than that measured in real experiments with NV centers. This is done to artificially reduce the integration time in numerical calculations and, thus, to considerably diminish computational efforts. This trick, however, causes no qualitative changes on the resulting scenarios presented below as the smallest time scale $\sim 1/\gamma_{\parallel}$ is well separated from the rest of the time scales.
 
We assume without loss of generality that the driving amplitude $\eta(t)$ is a real function. Taking into account that the $q$-Gaussian distribution is symmetric with respect to the mean frequency $\omega_s$ and that we always operate on resonance, $\omega_p=\omega_c=\omega_s$, it can straightforwardly be proven that the cavity amplitude $a(t)$ is real too.

\begin{figure}%[t!]
\includegraphics[angle=0,width=1.\columnwidth]{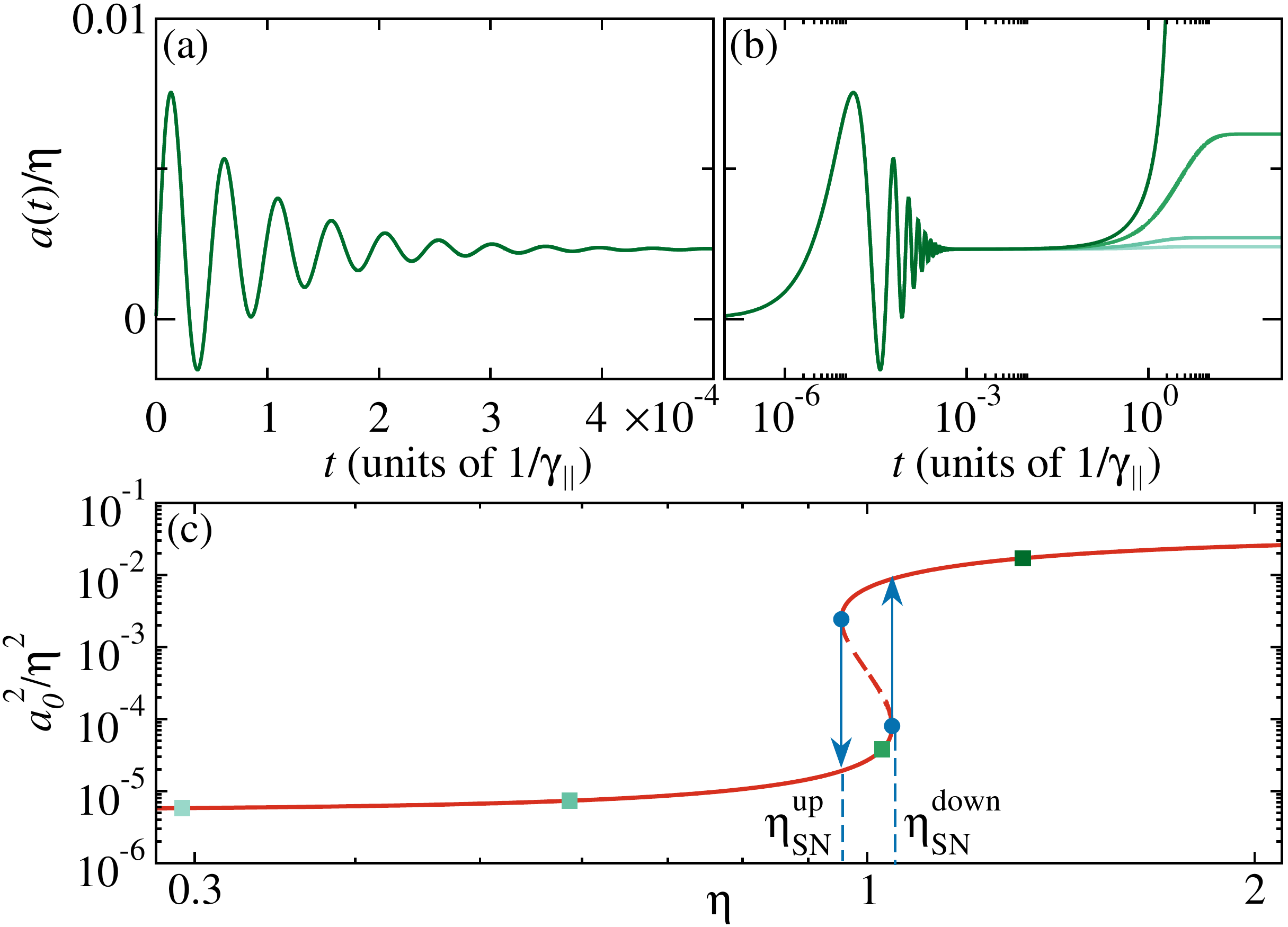}
%\vspace*{-0.6cm}
\caption{(a,b) Cavity probability amplitude $a(t)$ versus time $t$ (in units of $1/\gamma_{\parallel}$) under the action of an external drive with constant amplitude $\eta/\kappa=0.054, 0.11, 0.19, 0.24$ (from light to dark green), which is switched on at $t=0$. We present the numerical results on a linear time scale in (a) and on a logarithmic time scale in (b) to cover a much longer time interval. (c) Stationary solutions for the cavity amplitude $a_0^2$ as a function of the driving amplitude $\eta$ (log-log scale). {\it Solid and dashed curves} are stable and unstable solutions represented by a node and a saddle. The two points at which they meet are saddle-node bifurcations. Hysteresis behavior of $a_0^2$ under smooth sweeping of the amplitude $\eta$ through the critical region is indicated by two arrows. {\it Green symbols} are stationary solutions to which the system eventually settles for the values of $\eta$ from (a,b) [colours of the symbols in (c) correspond to those of the curves in (a,b)]. $\eta$ in (a-c) is normalized such that $\eta=1$ lies exactly at the middle between the upper and lower saddle-node (SN) bifurcations occurring at, respectively, $\eta_{SN}^{up} = 0.955$ and $\eta_{SN}^{down} = 1.045$ (solid circles).}
\label{fig_dynamic_bistab}
\end{figure}
%$\eta_{SN}^{up}/\kappa = 0.175$ and $\eta_{SN}^{down}/\kappa = 0.192$
% 

%
\section{Dynamics and bistability effect}
\label{Dynamics_bistab}

Typical temporal evolutions for the cavity probability amplitude $a(t)$ are shown in Fig.~\ref{fig_dynamic_bistab}(a,b) for the case when the external constant driving field $\eta$ is suddenly switched on at time $t=0$, i.e., $\eta(t)=\eta\, \Theta(t)$ with $\Theta(t)$ being the Heaviside step function and $\eta=\text{const}$. This choice for such a simple shape of the external drive allows us to capture all essential features of the dynamics from the trivial initial state to the final stationary state with comparatively simple equations. The linear and logarithmic time scales in Fig.~\ref{fig_dynamic_bistab}(a,b), respectively, show well-separated time scales of the resulting dynamics. A fast time scale, on which Rabi oscillations are clearly resolved, corresponds to the following inverse rates $\sim 1/\kappa, 1/\gamma_\perp$ [see Fig.~\ref{fig_dynamic_bistab}(a)]. Rabi oscillations are a signature that the system is in the strong-coupling regime due to the ensemble's strong collective coupling to the cavity. The decoherence caused mainly by inhomogeneous broadening of the spin ensemble, finally lets the oscillations disappear, giving rise to a transient steady state regime at rather long times. It is seen from Fig.~\ref{fig_dynamic_bistab}(a) that the dynamics is scalable over both of these time intervals as the ratio of the cavity amplitude to the amplitude of the driving signal, $a(t)/\eta$, remains practically unaltered even for moderate values of $\eta$ (all curves for different values of $\eta$ lie on top of each other in this figure). The situation is different, however, at the longest time scales in the system, of the order of  $\sim 1/\gamma_\parallel$. As shown in Fig.~\ref{fig_dynamic_bistab}(b), at such very long times, the value of $a(t)$ starts to deviate from a transient constant level and finally evolves to its ultimate stationary state having a strong dependence on the value of $\eta$. We will demonstrate explicitly below that under certain circumstances the system can evolve into a final stable state on time scales which are even much longer than $1/\gamma_\parallel$.

Next, we analyze stationary solutions obtained by setting all time derivatives in Eqs.~(\ref{Eq_a_Volt}-\ref{Eq_ck_Volt}) to zero so that a closed nonlinear equation with respect to the stationary cavity amplitude $a_0$ (indicated by the subscript 0) can be derived
\begin{eqnarray}
\label{Eqs_st_sol_a0}
a_0\left(1+\sum_k \dfrac{C_k}{1+a_0^2/n_k}\right)=\dfrac{\eta}{\kappa}.
\end{eqnarray}
Here $C_k$ and $n_k$ are, respectively, the cooperativity parameter and the photon saturation number for the $k$-th spin defined as
\begin{eqnarray}
\label{C_k_n_k_def}
C_k=\dfrac{g_k^2}{\gamma_\perp\kappa\left(1+\Delta_k^2/\gamma_\perp^2\right)},\,\,\,n_k=\dfrac{\gamma_\perp\gamma_\parallel}{4g_k^2}\left(1+\Delta_k^2/\gamma_\perp^2\right).\,\,\,\,\,\,\,\,\,
\end{eqnarray}
The collective system cooperativity can be defined as $C=\sum_k C_k$. The $z$-component of the $k$-th spin operator expectation value can then be calculated as follows
\begin{eqnarray}
\label{Eqs_st_sol_sz}
\sigma_k^{z0} = -\frac{1}{1+a_0^2/n_k}.
\end{eqnarray}

In Fig.~\ref{fig_dynamic_bistab}(c) all solutions of Eq.~(\ref{Eqs_st_sol_a0}) for $a_0^2$ are presented versus the driving amplitude $\eta$ for $\Omega/2\pi=12$~MHz. We choose a value for the collective cooperativity $C\approx 78$ which is larger than the threshold value $C_{th} \approx 42$ above which the system always features bistable behavior within a certain interval of $\eta$. The lower or cooperative branch in Fig.~\ref{fig_dynamic_bistab}(c) is characterized by rather small spin deviations from the south pole of the Bloch sphere and the enhanced cooperative emission resulting in low transmissions. Indeed, for small enough driving amplitudes $\eta$, the term $a_0^2/n_k$ can be neglected in Eqs.~(\ref{Eqs_st_sol_a0},\ref{Eqs_st_sol_sz}) giving rise to the pure linear response of the system with $\sigma_k^{z0}\approx -1$ and $a_0\approx\eta/[\kappa(1+C)]$, where the stationary amplitude is diminished by a factor of $1+C$ as compared to the amplitude $a_0=\eta/\kappa$ for the empty cavity without spin ensemble.

As displayed in Fig.~\ref{fig_dynamic_bistab}(c), stationary solutions lying on the lower branch represented by a node remains stable when the driving amplitude $\eta$ is below some critical value $\eta_{SN}^{down}$\!. At this critical value a stable node coalesces with another point lying on the unstable branch [dashed line in Fig. \ref{fig_dynamic_bistab}(c)] which is a saddle. Thus in dynamical system classification we are dealing here with a saddle-node (SN) bifurcation at $\eta=\eta_{SN}^{down}$ accompanied by a discontinuous transition to the upper branch, which is also represented by a stable node \cite{Glendinning}. If one starts from the upper branch and the driving amplitude $\eta$ is decreased, the system switches back to the lower branch at $\eta=\eta_{SN}^{up}$ where another SN bifurcation occurs. The upper branch is characterized by large spin deviations from the south pole leading to a progressive reduction of the spin-cavity coupling as $\eta$ increases. The limiting case of $\eta \rightarrow \infty$ is readily restored from Eqs.~(\ref{Eqs_st_sol_a0},\ref{Eqs_st_sol_sz}) and corresponds to the fully saturated spin ensemble, $\sigma_k^{z0}=0$, which is effectively decoupled from the cavity with $a_0=\eta/\kappa$.

The bistability region is located between the two critical points, $\eta=\eta_{SN}^{up}$ and $\eta=\eta_{SN}^{down}$, which can be determined from the condition that the derivative $d\eta/da_0$ vanishes (i.e. $da_0/d\eta$ is infinite - a signature of a first-order transition), 
\begin{eqnarray}
\label{Eq_deriv_deta_da0}
\dfrac{1}{\kappa}\dfrac{d\eta}{da_0}=1+\sum_k \dfrac{C_k}{(1+a_0^2/n_k)}\left(1-\dfrac{2a_0^2/n_k}{1+a_0^2/n_k}\right)=0.\,\,\,\,\,\,\,
\end{eqnarray}
Note that the unstable branch is accompanied by a negative slope of the driving strength $\eta$ as a function of the transmission amplitude $a_0$, $d\eta/da_0<0$, and its two ends connect the upper and lower branch for which $d\eta/da_0>0$ holds. It is worth noting that Eq.~(\ref{Eq_deriv_deta_da0}) has a solution only when the cooperativity parameter $C$ is above a certain threshold value $C_{th}$ -- otherwise the bistability effect does not occur. In the latter case the system features no phase transition when starting from the lower branch and increasing $\eta$. Rather, the lower and upper branches are directly connected with each other at the point where $d^2a_0/d\eta^2=0$.

\begin{figure}
\includegraphics[angle=0,width=1.\columnwidth]{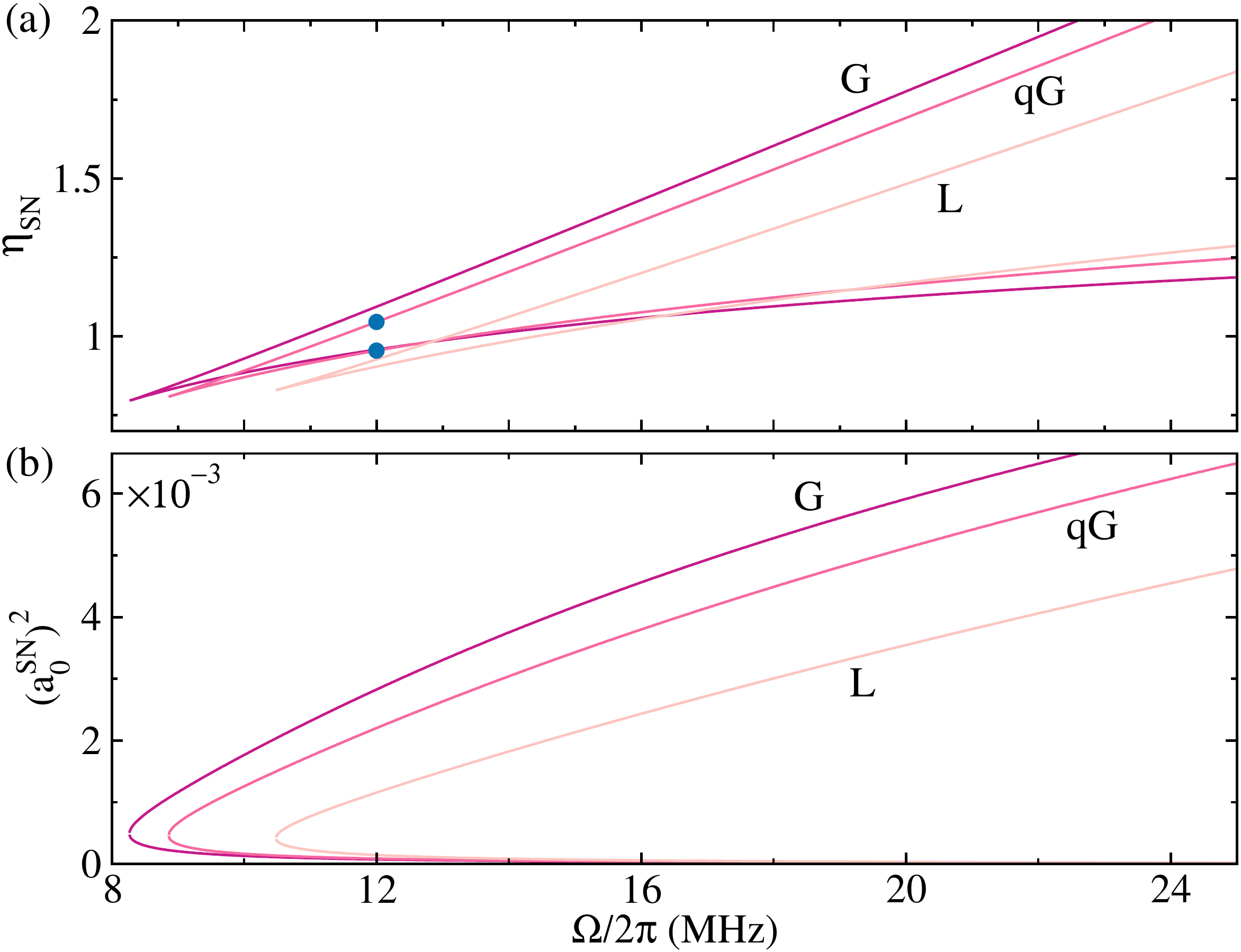}
%\vspace*{-0.6cm}
\caption{Pairs of critical values for the driving amplitudes $\eta_{SN}$ (a) and corresponding cavity amplitudes squared $(a_0^{SN})^2$ (b) at which SN bifurcations occur for Gaussian, $q$-Gaussian and Lorentzian distributions (labeled by G, qG and L in both panels). Solid circles in (a) are the upper and lower SN bifurcations from Fig.~\ref{fig_dynamic_bistab} and for the $q$-Gaussian distribution the same parameters are used as in Fig.~\ref{fig_dynamic_bistab}.}
\label{fig_SN_eta_vs_Omega}
\end{figure}

We now explore the onset of bistability for different shapes of the spectral spin distribution $\rho(\omega)$. The results of the corresponding calculations are displayed in Fig.~\ref{fig_SN_eta_vs_Omega} for a Gaussian, $q$-Gaussian and a Lorentzian distribution as a function of the collective coupling strength $\Omega$. (Note that the value of cooperativity $C$ monotonically grows with $\Omega$.) One can see from this figure that the onset of bistability has a general tendency to move towards higher values of the coupling strength $\Omega$ as the distribution becomes broader. This can be explained by the fact that for broader distributions spectrally more distant spins are effectively less and less coupled to a cavity, and as a result, larger values for the critical coupling strength are required to observe the onset of bistability. 

Another interesting observation is that critical values for the driving amplitudes $\eta_{SN}$ as well as resulting values for the cavity amplitudes $a_0^{SN}$ are almost independent of the shape of the spectral spin distribution at the onset of bistability (see cusps in Fig.~\ref{fig_SN_eta_vs_Omega}(a,b) from which a pair of curves emanate). This can be intuitively understood as follows: regardless of the precise shape of $\rho(\omega)$ a certain value of nonlinearity is needed to trigger the instability mechanism leading to a discontinuous transition. As the value of nonlinearity is directly related to the collective characteristic quantity $|a_0|^2$, the only crucial issue is at which critical value for the coupling strength $\Omega$ the necessary value for the onset of bistability $|a_0^{SN}|^2$ is achieved. We also note that, as the coupling strength $\Omega$ increases, the bistability region gets wider in $\eta$ more rapidly for narrower shapes of the spectral density.

\section{Adiabatic elimination and critical slowing-down}
\label{Sec_Adiab_elim}
\begin{figure}%[t!]
\includegraphics[angle=0,width=1.\columnwidth]{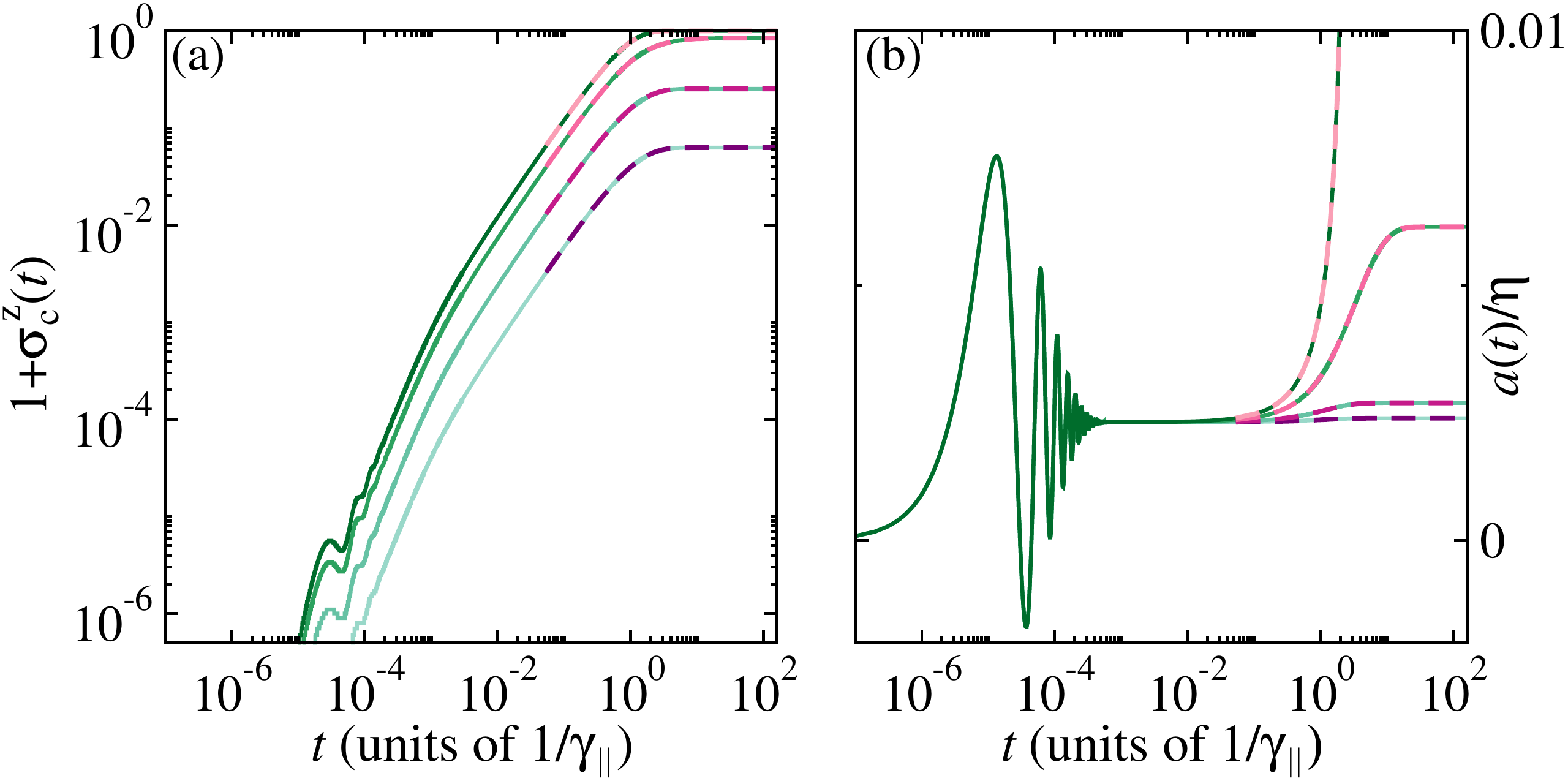}
%\vspace*{-0.6cm}
\caption{The $z$-component of the central spin operator expectation value, $\sigma_c^z(t)$, [(a) in log-log scale] and the cavity probability amplitude $a(t)$ [(b) in log-lin scale] versus time $t$ (in units of $1/\gamma_{\parallel}$) under the action of an external drive with constant amplitude, $\eta(t)=\eta\, \Theta(t)$, where $\Theta(t)$ is the Heaviside step function and $\eta=\text{const}$. The values for $\eta$ are chosen to be the same as those shown by symbols in Fig.~\ref{fig_dynamic_bistab}(c). {\it Solid lines}: full numerical solutions. {\it Dashed lines}: numerical solution of Eqs.~(\ref{Eq_sig_z_adiab}, \ref{Eq_a_adiab}) obtained under adiabatic elimination of $\sigma_k^-(t)$ and $a(t)$.}
\label{fig_adiab_elimination}
\end{figure}

To get more detailed insights into the system dynamics, we use the standard procedure of adiabatic elimination of selected dynamical variables \cite{Abraham:1980aa,Drummond:1981ab,Lugiato:1984aa}. Taking into account that $\gamma_\parallel \ll \kappa,\gamma_{\perp}, \Omega$, we expect that all long-lived processes occur on time scales of the order of $1/\gamma_\parallel$. Specifically, we adiabatically eliminate the cavity amplitude $a$ and the spin lowering expectation value $\sigma_k^-$ from Eqs.~(\ref{Eq_a_Volt}-\ref{Eq_ck_Volt}), as these expectation values adiabatically follow the evolution of the $z$-component of the spin operator expectation value $ \sigma_k^z$ at large times when $t \gg 1/\kappa,1/\gamma_{\perp}, 2\pi/\Omega$. We first introduce the dimensionless time, $\tau=\gamma_\parallel t$, and rewrite  Eqs.~(\ref{Eq_a_Volt}-\ref{Eq_ck_Volt}) as follows
\begin{subequations}
\begin{eqnarray}
\label{Eq_a_Volt_adiab}
&&\dfrac{\gamma_\parallel}{\kappa}\dfrac{da}{d\tau}= -a - \dfrac{1}{2\kappa}\sum\nolimits_k g_k  \sigma_k^y+\dfrac{\eta}{\kappa}, 
\\[6pt]
\label{Eq_bk_x_Volt_adiab}
&&\dfrac{\gamma_\parallel}{\gamma_\perp}\dfrac{d \sigma_k^x}{d\tau} = - \sigma_k^x-\dfrac{\Delta_k}{\gamma_\perp}\sigma_k^y,
\\[6pt]
\label{Eq_bk_y_Volt_adiab}
&&\dfrac{\gamma_\parallel}{\gamma_\perp}\dfrac{d \sigma_k^y}{d\tau} =\dfrac{\Delta_k}{\gamma_\perp}\sigma_k^x-\sigma_k^y-\dfrac{2g_k}{\gamma_\perp} \sigma_k^z a,
\\[6pt]
\label{Eq_ck_Volt_adiab}
&&\dfrac{d \sigma_k^z}{d\tau} = -(1+ \sigma_k^z)+\dfrac{2g_k}{\gamma_{\parallel}}\sigma_k^y a,
\end{eqnarray}
\end{subequations}
where $\sigma_k^-=(\sigma_k^x-i \sigma_k^y)/2$. Note that in Eq.~(\ref{Eq_a_Volt_adiab}) we have used the fact that the $x$-component of the collective spin, $J_x=\sum\nolimits_k g_k  \sigma_k^x/2$, vanishes since spectral densities under consideration are always symmetric with respect to the central spin frequency $\omega_s$ and $\omega_c=\omega_s$.

Next, from the inequalities, $\gamma_\parallel/\kappa \ll 1$ and $\gamma_\parallel/\gamma_\perp \ll 1$, we infer that the time derivatives of the variables to be eliminated, $d\sigma_k^-/d\tau$ and $da/d\tau$, give negligibly small contributions at large times with respect to the terms staying on the right-hand side of Eqs.~(\ref{Eq_a_Volt_adiab}-\ref{Eq_bk_y_Volt_adiab}). Straightforward calculations then yield the following reduced equations for $\sigma_k^z$ and $a$: 
\begin{eqnarray}
\label{Eq_sig_z_adiab}
&&\dfrac{d\sigma_k^z}{d\tau}=-(1+\sigma_k^z)-\dfrac{\eta^2}{\kappa^2}\dfrac{\sigma_k^z}{n_k\left(1-\sum_l C_l\sigma_l^{z}\right)^2},
\\[6pt]
\label{Eq_a_adiab}
&&a=\dfrac{\eta}{\kappa\left(1-\sum_lC_l\sigma_l^z\right)},
\end{eqnarray}
where the cooperativity parameter $C_l$ and the photon saturation number $n_l$ for the $l$-th spin are given by Eq.~(\ref{C_k_n_k_def}). Note that further simplification of these equations without any additional assumptions regarding the shape of inhomogeneous broadening turns out to be difficult. In particular, we could not derive a closed equation with respect to the cavity amplitude $a$. 

As can be deduced from the derivation of the adiabatic elimination, the above equations are of restricted validity in the sense that they can not capture the effect of initial coherent energy exchange (Rabi oscillations) between the cavity and the spin ensemble, but will instead describe the evolution at large times only. Indeed, the cavity amplitude $a$ is enslaved to $\sigma_k^z$ through Eq.~(\ref{Eq_a_adiab}): at every instant of (slow) time the value of $a$ is entirely determined by the spin components $\sigma_k^z$, which are given by the closed set of Eqs.~(\ref{Eq_sig_z_adiab}). Note also that solving the equations after adiabatic elimination in general needs special caution in the choice of system parameters and the time step of numerical integration: There is a number of additional requirements to be simultaneously fulfilled for the global validity of the adiabatic approximation, besides a well-defined time scale separation (such as the magnitude of all parameters, of the physical variables, and of their fluctuations, see \cite{Lugiato:1984aa} for details). 

In Fig.~\ref{fig_adiab_elimination}, the results of the calculations under adiabatic elimination are compared with those obtained in the framework of the full Maxwell-Bloch equations (\ref{Eq_a_Volt}-\ref{Eq_ck_Volt}). In the former case Eqs.~(\ref{Eq_sig_z_adiab}) for $\sigma_k^z$ are numerically solved with initial conditions $\sigma_k^z=-1$ (spin ensemble is in the ground state) and $a$ is correspondingly found from Eq.~(\ref{Eq_a_adiab}) (enslaved variable). Thus, after performing the adiabatic elimination all details about initial cavity population and its subsequent transient dynamics are completely washed out. It is seen from Fig.~\ref{fig_adiab_elimination} that the adiabatic elimination indeed is a very reasonable approximation for the system's evolution at large times after the transient oscillatory behavior disappears. 

Although $a$ varies slowly in time and the derivative $da/d\tau$ was omitted in Eq.~(\ref{Eq_a_Volt_adiab}) owing to the small prefactor as mentioned above, we can still capture this slow variation of the cavity amplitude $a$ by differentiating the reduced Eq.~(\ref{Eq_a_adiab}) with respect to slow time $\tau$. We finally arrive at the most general expression for $da/d\tau$ 
\begin{eqnarray}
\label{Eq_a_adiab_diff}
\dfrac{da}{d\tau}&=&\dfrac{\kappa a^2}{\eta}\sum_l C_l\, \dfrac{d\sigma_l^z}{d\tau}, 
\end{eqnarray}
where $d\sigma_l^z/d\tau$ is determined by Eq.~(\ref{Eq_sig_z_adiab}).

If we now assume that all spins are in resonance, $\Delta_k=0$ and $\sigma_l^z=\sigma^z$,  Eq.~(\ref{Eq_a_adiab_diff}) can be drastically simplified so that we obtain a single ODE for the cavity amplitude \cite{Angerer:2017aa}:
\begin{eqnarray}
\label{Eq_a_nonlin_all_reson}
\dfrac{da}{d\tau}=a-\dfrac{\kappa}{\eta}(1+C)a^2+\dfrac{4\kappa C}{N\gamma_{\parallel}}a^3-\dfrac{4\kappa^2C}{N\gamma_{\parallel}\eta}a^4,
\end{eqnarray}
where $C=g^2 N/(\gamma_{\perp}\kappa)$ stands for the collective system cooperativity. Thus a polynomial structure of this ODE with respect to $a$ in the absence of inhomogeneous broadening acquires a more complex form than the simplest normal form of a differential equation exhibiting a saddle-node (SN) bifurcation \cite{Strogatz00, expl_norm_form}. Note that exactly at the SN bifurcation two stationary solutions (one stable and one unstable fixed point) coalesce with each other being solutions of the nonlinear algebraic equation, $\dot{x}=f(x)=0$ (the dot stands for the time derivative). When slightly detuning the control parameter from the threshold for the SN bifurcation to the side where the above mentioned equation has no solutions, the value of $\dot{x}$ may be arbitrary small.  As a result, the system passes very slowly through the region in phase space $(x,\dot{x})$ near the SN bifurcation. In the literature this phenomenon is often referred to as a saddle-node ``ghost'' \cite{Strogatz00} since the phase trajectories are considerably delayed in their flow by the SN bifurcation before they eventually reach a corresponding stable solution (the so-called critical slowing-down effect). Moreover a set of trajectories exhibits a dip close to the SN bifurcation so that the formed structure in phase space is reminiscent of a ``bottleneck''  into which they are funneled.

%Moreover the structure in phase space formed by a set of trajectories close to the SN bifurcation and is reminiscent of a ``bottleneck'' structure.
%It is well known from the literature \cite{Strogatz00} that a saddle-node ``ghost'' arises when slightly mismatching from the threshold for the SN bifurcation at which two fixed points coalesce resulting in a slow passage through a region which is reminiscent of a ``bottleneck'' structure in the phase space (the so-called critical slowing-down effect). 

%
\begin{figure}%[t!]
\includegraphics[angle=0,width=1.\columnwidth]{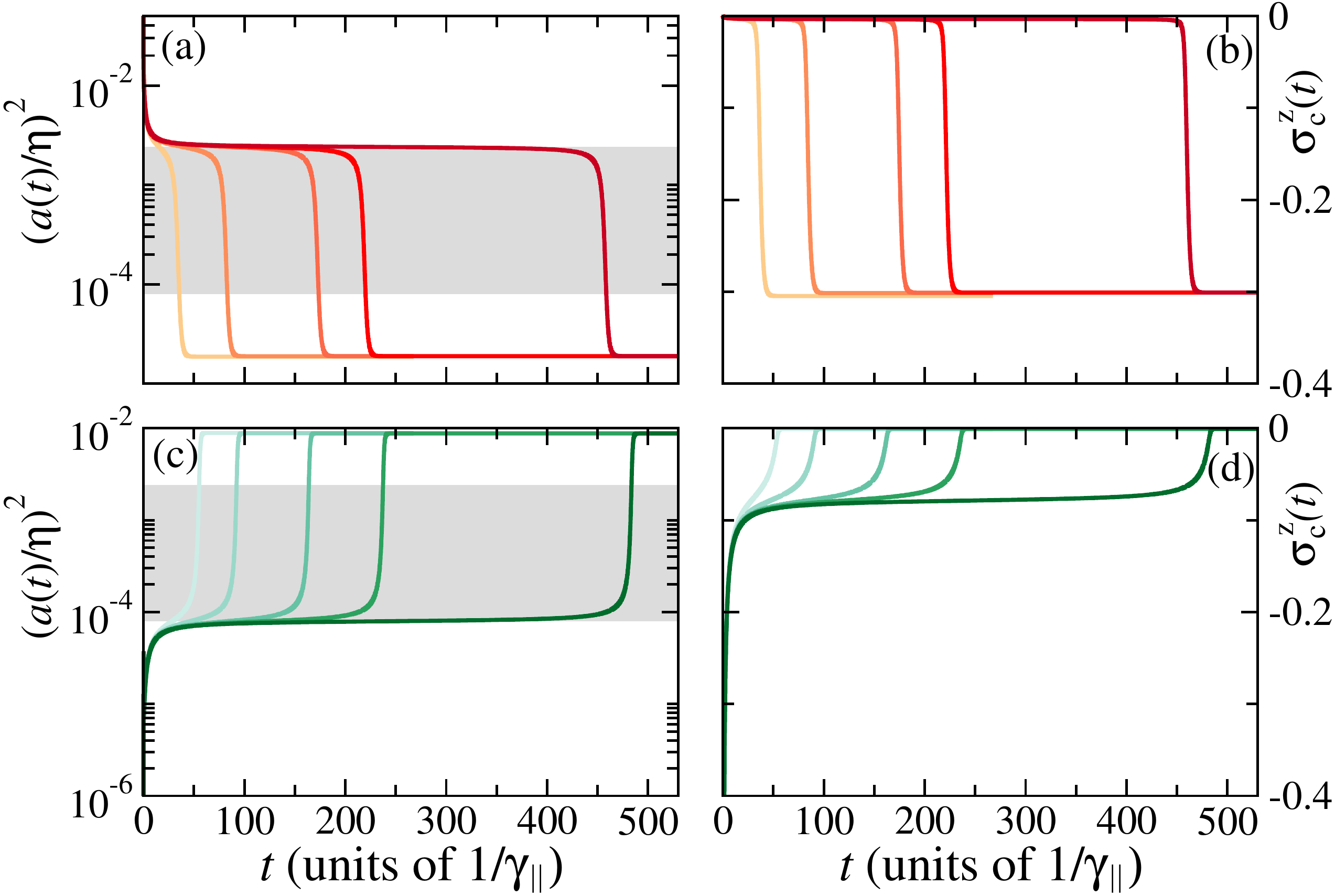}
%\vspºace*{-0.6cm}
\caption{Long transient quench dynamics in the vicinity of the upper (a), (b) and lower (c), (d) SN bifurcations shown in Fig.~\ref{fig_dynamic_bistab}. Cavity probability amplitude squared $a^2(t)$ [left panels] and the $z$ component of the central spin operator expectation value, $\sigma_c^z(t)$ [right panels], versus time $t$ (in units of $1/\gamma_{\parallel}$) under the action of an external drive with constant amplitude, $\eta(t)=\eta\, \Theta(t)$, where $\Theta(t)$ is the Heaviside step function and $\eta=\text{const}$. The values for $\eta$ are chosen slightly below (above) the upper (lower) SN bifurcation. As an initial condition a stationary solution on the upper branch [(a), (b)] or the lower branch [(c), (d)] is chosen,  which lies far away from the critical region. The closer the value of $\eta$ to the corresponding critical value $\eta_{SN}^{up}$ or $\eta_{SN}^{down}$, the longer the transient time (transient time increases from {\it light} to {\it dark curves} in all panels). Gray regions designate a gap in values of $|a_0|^2$, where no stable stationary solution exists (see Fig.~\ref{fig_dynamic_bistab}).}
\label{fig_slow_dynamics_NEW_III}
\end{figure}

Based on this simple case, we conjecture that such phenomena as the saddle-node ghost and the critical slowing-down effect can also show up for the case with inhomogeneous broadening governed by the full Maxwell-Bloch equations (\ref{Eq_a_Volt}-\ref{Eq_ck_Volt}) or their reduced version after adiabatic elimination [see Eqs.~(\ref{Eq_sig_z_adiab}-\ref{Eq_a_adiab})], despite the fact that one cannot derive a single differential equation of polynomial form for the cavity amplitude $a$. In Fig.~\ref{fig_slow_dynamics_NEW_III} we present one specific example of such effects, i.e., the quench dynamics in the vicinity of the upper and lower SN bifurcations from Fig.~\ref{fig_dynamic_bistab}. Specifically, we take as initial conditions for $a$ and $\sigma_k^z$ the stationary solution located on the upper branch at a certain value for the driving amplitude $\eta$, which is substantially larger than the critical value $\eta_{SN}^{up}$ at which the upper SN bifurcation occurs. Next, the system is quenched below the upper SN by suddenly changing $\eta$ to the values which lie slightly below $\eta_{SN}^{up}$. A similar test is also repeated for the lower SN bifurcation but now we start from a stationary solution on the lower branch and abruptly change $\eta$ to the values slightly above $\eta_{SN}^{down}$. In both cases we observe a very pronounced bottleneck structure developing over time scales much longer than the slowest time scale in our system, $1/\gamma_\parallel$. This behavior is indicative of the aforementioned critical slowing-down phenomenon in systems exhibiting saddle-node bifurcations \cite{Kuehn09,Bonifacio:1979aa,Barbarino:1982aa}.

\begin{figure}
\includegraphics[angle=0,width=1.\columnwidth]{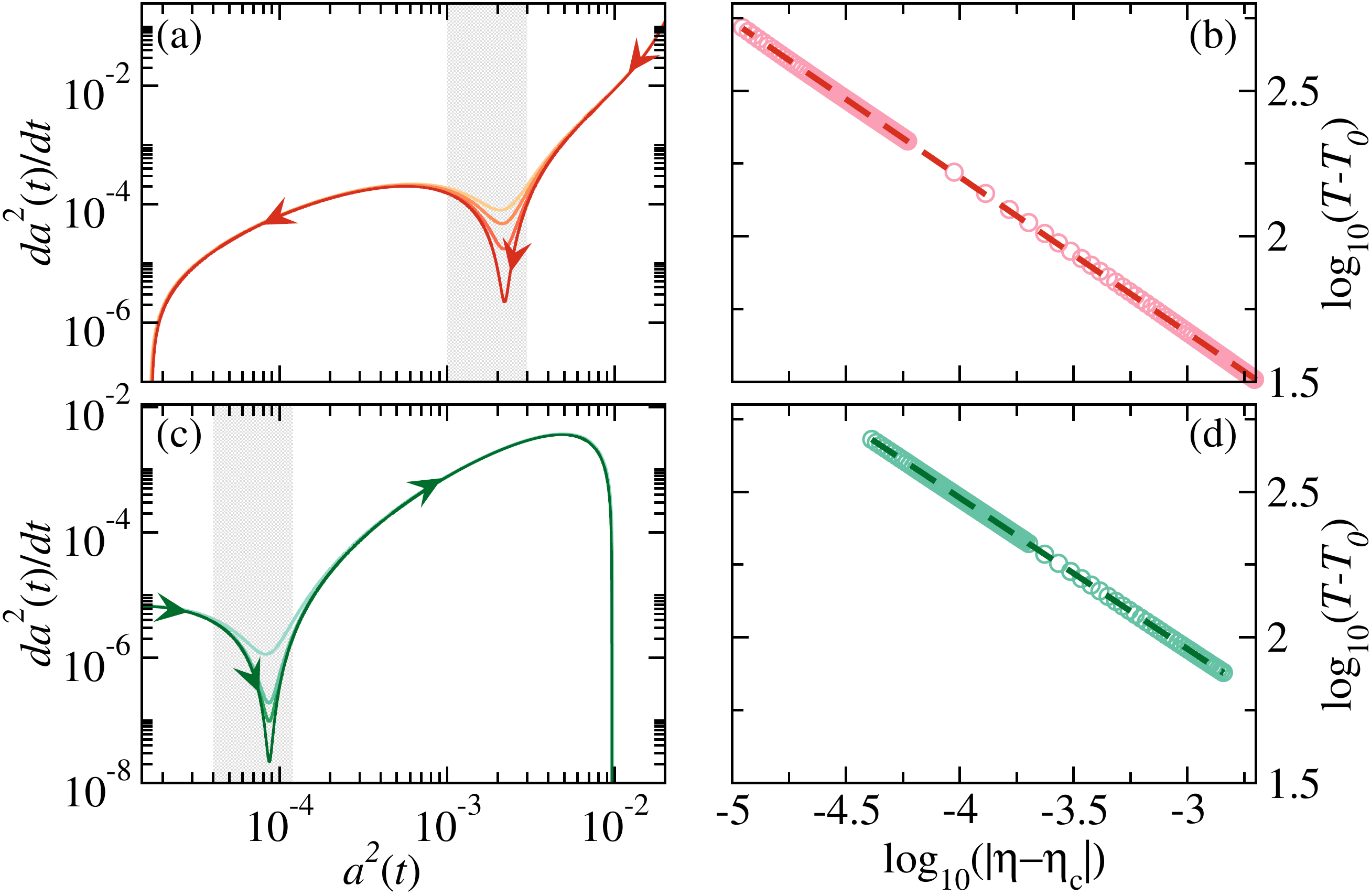}
%\vspace*{-0.6cm}
\caption{Phase portraits and scalings for the quench dynamics near SN bifurcations. (a) Phase trajectories in the $(da^2(t)/dt,a^2(t))$ plane (log-log scale) for an initial condition chosen as a stationary state lying on the upper branch of the S-shaped bistability curve in Fig.~\ref{fig_dynamic_bistab}(c). The driving value $\eta$ is then abruptly decreased to a value located slightly below the one at which the upper SN bifurcation occurs, $\eta_c \equiv \eta_{SN}^{up} = 0.955$ (in the normalization of Fig.~\ref{fig_dynamic_bistab}). For the purpose of demonstration a few values in the proximity of the bifurcation point are taken. (b) The time, $T-T_0$ (measured in units of $1/\gamma_{\parallel}$), needed to pass the slowing-down region [{\it gray area} in (a)] versus $\log_{10}(|\eta-\eta_c|)$ is displayed by {\it empty symbols}. {\it Dashed curve}: the algebraic fit, $T-T_0 = \beta\,|\eta-\eta_c|^{-\alpha}$, with the exponent $\alpha=0.53$, $T_0=-6.25$ and $\beta=1.16$. (c), (d) Corresponding vizualisations of the quench dynamics near the lower SN bifurcation at $\eta_c \equiv \eta_{SN}^{down}=1.045$. A stationary solution located at the lower branch is now used as an initial condition and $\eta$ is increased to a value located slightly above this bifurcation point. {\it Dashed curve}: the same algebraic fit as above with the exponent $\alpha=0.52$, $T_0=-74.7$ and $\beta=2.54$. Arrows in (a), (c) indicate the system's evolution during the course of time. The darker a {\it curve's color} in (a), (c), the closer the value of $\eta$ to the corresponding critical value $\eta_{SN}^{up}$ or $\eta_{SN}^{down}$.}
\label{fig_TRAJECT_NEW_UP_LOW}
\end{figure}
%

%Long transient quench dynamics in the vicinity of the upper (a), (b) and lower (c), (d) SN bifurcations shown in Fig.~\ref{fig_dynamic_bistab}. Cavity probability amplitude squared $|a(t)|^2$ [left panels] and the $z$ component of the central spin operator expectation value, $\sigma_c^z(t)$ [right panels], versus time $t$ (in units of $1/\gamma_{\parallel}$) under the action of an incident pulse with constant amplitude, $\eta(t)=\eta\, \Theta(t)$, where $\Theta(t)$ is the Heaviside step function and $\eta=\text{const}$. The values for $\eta$ are chosen slightly below (above) the upper (lower) SN bifurcation. As an initial condition a stationary solution on the upper branch [(a), (b)] or the lower branch [(c), (d)] is chosen,  which lies far away from the critical region. The closer the value of $\eta$ to the corresponding critical value $\eta_{SN}^{up}$ or $\eta_{SN}^{down}$, the longer the transient time (transient time increases from {\it light} to {\it dark curves} in all panels). Gray regions designate a gap in values of $|a_0|^2$, where no stable stationary solution exists (see Fig.~\ref{fig_dynamic_bistab}). 

In order to further characterize the slowing-down dynamics, we plot in Fig.~\ref{fig_TRAJECT_NEW_UP_LOW}(a,\,c) several phase trajectories. The closer the driving amplitude $\eta$ is to the critical value at which the upper or lower SN bifurcation occurs, the more and more time our system spends near a narrow slowing-down region in phase space [gray areas in Fig.~\ref{fig_TRAJECT_NEW_UP_LOW}(a,c)], where $da^2\!/dt$ drops to very small values (see a distinct dip structure in the shape of phase trajectories). At the critical point a stable node and a saddle collide with each other resulting in the divergence of time $T$, which is the time a phase trajectory spends in the slowing-down region displayed by gray areas in Fig.~\ref{fig_TRAJECT_NEW_UP_LOW}(a,c). Such a singular behavior is explained by a vanishing ``velocity'' $da^2\!/dt$ exactly at the SN bifurcation. We found that this divergence has an algebraic nature, by fitting the calculated values for $T$ to the function $T=T_0 + \beta\,|\eta-\eta_c|^{-\alpha}$, where $\eta_c=\eta_{SN}^{up}$ or $\eta_c=\eta_{SN}^{down}$ for the upper or lower SN bifurcation, respectively [see Fig.~\ref{fig_TRAJECT_NEW_UP_LOW}(b,\,d)]. 

It turns out that in both cases the exponent $\alpha$ only slightly exceeds the well-known square-root scaling law, $\alpha=0.5$, for the simplest normal form of a non-degenerate SN bifurcation, $d x/dt=r+x^2$, where $x\in Re$ and $r\le0$ is the bifurcation parameter \cite{Kuehn09}. Such scaling similarities can be traced back to very generic features of continuous phase transitions at which a system becomes scale-invariant and is characterized by an infinite correlation length and time. Specifically, both correlation length and time demonstrate power law divergence upon changing the external parameter in the vicinity of the phase transition \cite{Vojta:2003aa}. Moreover, a set of critical exponents can be the same for a certain class of phase transitions which share the same symmetries and dimensionality. This phenomenon referred to as ``universality'' \cite{Vojta:2003aa} can be understood by divergent correlations at the phase transition giving rise to smearing out of the system's complexity nearby it. Therefore, a very complex system can respond similarly as a very simple one provided that both are sufficiently close to the phase transition - a scenario which is realized in our case as well.

\section{Asymptotic decay}
\label{Sec_asympt_decay}

Regardless of the amount of time that should elapse to escape from the slowing-down region [see, e.g., long transient  plateaux in Fig.~\ref{fig_slow_dynamics_NEW_III} and gray areas in Figs.~\ref{fig_TRAJECT_NEW_UP_LOW}(a),\,(c)], a phase trajectory ultimately passes through this region. Finally, our system evolves towards a single possible stable state at the corresponding value of $\eta$ chosen during the quench procedure, see very left and right parts of Fig.~\ref{fig_TRAJECT_NEW_UP_LOW}(a) and (c). (Recall that the values of driving amplitude $\eta$ lie slightly below the critical value for the upper SN bifurcation, $\eta<\eta_{SN}^{up}$, and slightly above the one for the lower SN bifurcation, $\eta>\eta_{SN}^{down}$.) In both cases the stable solution is represented by a stable node. 

When approaching this final stable state, the dynamics again significantly slows down as $da^2\!/dt$ gets smaller and smaller during the course of time [see very left and right parts of Fig.~\ref{fig_TRAJECT_NEW_UP_LOW}(a),\,(c), respectively]. We now aim at finding the decay rate $\zeta$ of this asymptotic dynamics by a linear stability analysis around the stationary state for which all previous transient evolutions are irrelevant. Specifically, we slightly perturb the stationary states by writing
\begin{eqnarray}
\label{Ansatz_spin_cav_mod}
a&=&a_0+\delta a(t),\,\sigma^{\pm}_k=\sigma^{- 0}_k+\delta \sigma^{-}_k(t), 
\\[6pt]\nonumber
\sigma^z_{k}&=&
\sigma^{z0}_k+\delta \sigma^z_{k}(t),
\end{eqnarray}
where $a_0$, $\sigma^{-0}_k$ and $\sigma^{z0}_k$ are stationary solutions given by Eqs.~(\ref{Eqs_st_sol_a0},\ref{Eqs_st_sol_sz}) and $\delta a(t),\,\delta \sigma^{-}_k(t), \,\delta \sigma^z_{k}(t) \sim e^{-\zeta t}$ describe small perturbations around them. By substituting Eq.~(\ref{Ansatz_spin_cav_mod}) into the dynamical equations (\ref{Eq_a_Volt}-\ref{Eq_ck_Volt}) and neglecting the terms higher than first order, we end up after some algebra with the following charactersitic equation with respect to the decay rate $\zeta$
\begin{eqnarray}
\label{Eq_transc_zeta}
&&\zeta-\kappa-
\\[6pt]\nonumber
&&\sum_k\dfrac{g_k^2\sigma_k^{z0}(\zeta-\gamma_{\parallel}+4 \kappa |a_0|^2 C_k) (\zeta-\gamma_{\perp})}{((\zeta-\gamma_{\perp})^2+\Delta_k^2)(\zeta-\gamma_{\parallel})+4g_k^2 |a_0|^2(\zeta-\gamma_{\perp})}=0,
\end{eqnarray}
where we use the same notations as before. 

\begin{figure}
\includegraphics[angle=0,width=1.\columnwidth]{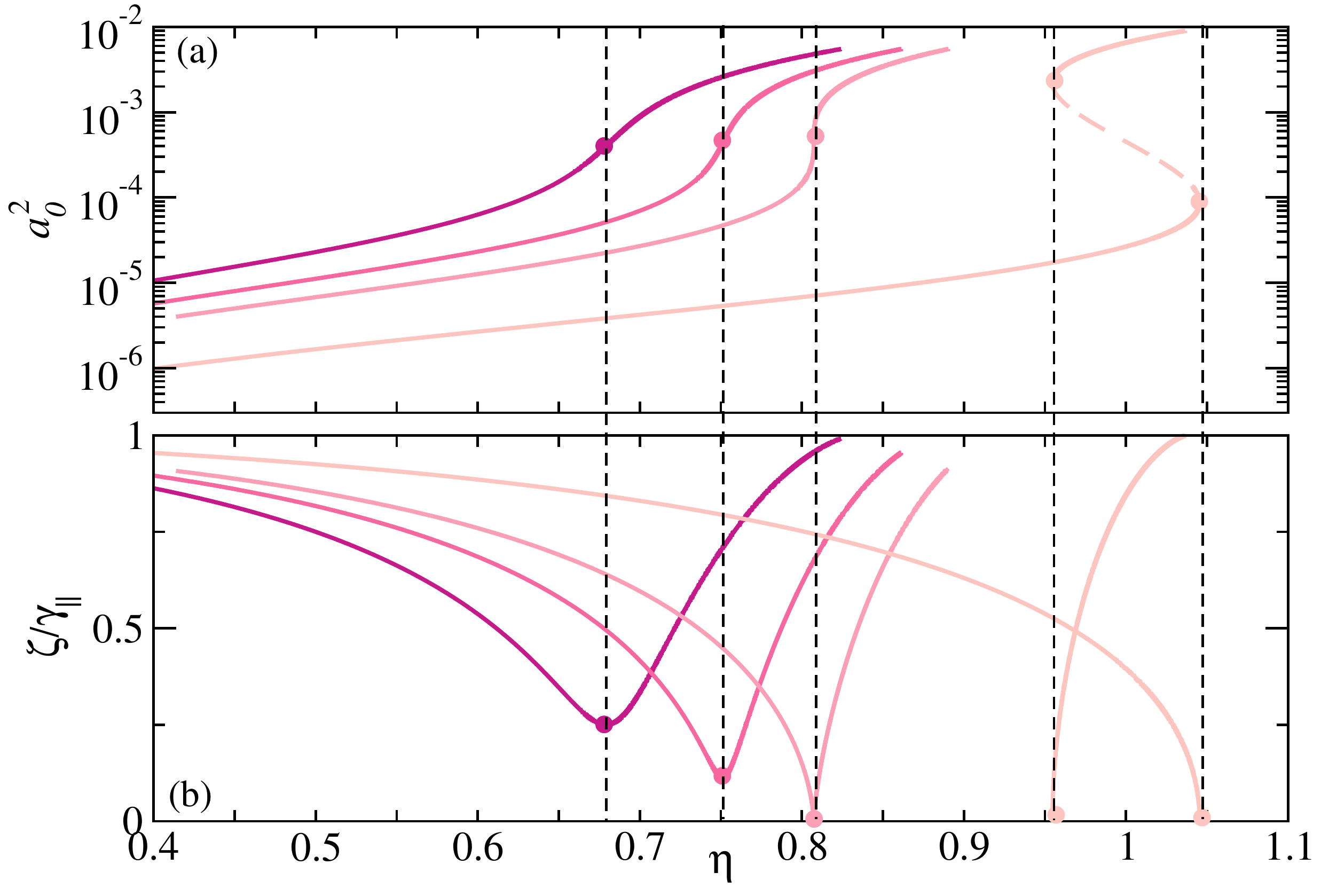}
%\vspace*{-0.6cm}
\caption{{\it Upper panel (a)}: Stationary solutions for the cavity amplitude $a_0^2$ versus driving amplitude $\eta$ for different values of the coupling strength, $\Omega/2\pi=7,8,8.84,12$~MHz (curves from left to right). The very right curve coincides with the one depicted in Fig.~\ref{fig_dynamic_bistab}(c). {\it Lower panel (b)}: The decay rate $\zeta$ (in units of $\gamma_{\parallel}$) towards the corresponding stationary states displayed in the upper panel (see Eq.~(\ref{Eq_transc_zeta})). {\it Symbols} on the {\it lower panel} are minima of $\zeta$ connected with the corresponding stationary states (symbols on the {\it upper panel}) by {\it dashed lines}. $\eta$ is normalized as in Fig.~\ref{fig_dynamic_bistab}.}
\label{fig_asymptotic_decay}
\end{figure}

We solve this equation numerically using a standard Newton-Raphson method and present the results for the lowest positive value of $\zeta$ for different values of the coupling strengths $\Omega$ in Fig.~\ref{fig_asymptotic_decay}. While nonlinear equations typically give rise to multiple solutions for $\zeta$, here the contributions from all values of $\zeta$ exponentially fade out as $e^{-\zeta t}$ and the one with the smallest value will eventually survive in the asymptotic limit ($t\rightarrow \infty$). 

As seen from Fig.~\ref{fig_asymptotic_decay}(b), the calculated curves for the decay rate $\zeta$ demonstrate a pronounced dip structure, which becomes steeper and steeper as one approaches the onset value for the SN bifurcation, $\Omega_{th}/2\pi=8.86$~MHz which corresponds to the collective cooperativity $C_{th} \approx 42$ (see three sets of curves in Fig.~\ref{fig_asymptotic_decay} for $\Omega/2\pi=7,8$ and $8.84$~MHz). Note, that in the whole range of $\eta$ displayed in Fig.~\ref{fig_asymptotic_decay}(b) the values for the decay rate $\zeta$ are smaller than the slowest longitudinal relaxation rate $\gamma_\parallel$ ($\zeta/\gamma_\parallel<1$ in this figure). Remarkably, the minimal value of $\zeta$ vanishes precisely at the SN bifurcation, resulting in the divergence of the corresponding time scale $1/\zeta$. As previously discussed, for even larger coupling strength ($\Omega/2\pi>8.86$~MHz) a pair of SN bifurcations occurs, giving rise to two separate curves for the values of $\zeta$ (see the disconnected curves for $\Omega/2\pi=12$~MHz in Fig.~\ref{fig_asymptotic_decay}). Each of the two different values for $\zeta$ in the bistability region stands for the decay rate with which a system exponentially decays in the limit of $t\rightarrow\infty$ to the corresponding stable stationary state lying either on the upper or on the lower branch. At both SN bifurcations $1/\zeta$ diverges as in the aforementioned case of a single SN bifurcation. 

We can now establish an interesting connection of the slow asymptotic dynamics explored above, which has an entirely classical nature, with the behavior in open quantum systems exhibiting dissipative phase transitions \cite{Carmichael:2015aa,Kessler:2012aa}. In quantum systems with a Markovian bath the dynamics is governed by a Lindblad master equation for the system's density matrix, which can be formulated in terms of a Liouvillian superoperator. The properties of its eigenvalue spectrum determine the resulting dynamics. Note that eigenvalues of the Liouvillian supermatrix are complex-valued among which a zero eigenvalue is always present. A usually non-degenerate state which belongs to this zero eigenvalue is associated with the unique quantum stationary state. It turns out that in different quantum systems, e.g., in a single-mode cavity with a Kerr optical nonlinearity driven by a laser, the next eigenvalue with the smallest imaginary part can come very close to zero in a certain range of control parameters, whereas its real part is identically zero therein \cite{Casteels:2016aa,Rodriguez:2017aa}. In this case the imaginary part of this eigenvalue represents the asymptotic relaxation rate to the quantum stationary state whose value might be much smaller than all other relaxation rates in the system - a phenomenon which is often referred to as a closure of the Liouvillian gap \cite{Casteels:2016aa,Rodriguez:2017aa, Minganti:2018aa,Reimer:2018aa,krimer:2018ac}. The structure of the Liouvillian gap closure as a function of the control parameter $\eta$ is reminiscent of the parameter dependence of the slowest decay rate $\zeta$ for the coupling strengths $\Omega$ below the onset of the SN bifurcation [see three shapes of $\zeta$ for $\Omega/2\pi=7,8,8.84,12$~MHz in Fig.~\ref{fig_asymptotic_decay}(b)] suggesting an interesting semiclassical-to-quantum similarity.

\section{Conclusions}
\label{Sec_Concl}
We have theoretically studied driven dissipative nonlinear dynamics and phase transitions for a single cavity mode interacting with an inhomogeneously broadened spin ensemble. We considered the thermodynamic limit treating the problem in the framework of Maxwell-Bloch equations assuming different shapes for the spectral spin density. Our main focus was concentrated on spin-cavity systems where the cavity decay rate and the non-radiative dephasing of individual spins are much larger than their radiative decay. We analyzed in detail the onset of bistability for different shapes of the spectral spin density and found bistability to always arise provided that the collective coupling strength is above a certain value. The dynamics under the action of a constant drive that is suddenly switched on turned out to be linear (and therefore scalable) on time scales featuring damped Rabi oscillations followed by a transient stationary state even for moderate amplitudes of the driving signal. In contrast, at times which are of the order of the slowest time scale proportional to the inverse of the radiative dephasing rate, the dynamics is nonlinear being perfectly captured by a much simpler set of equations obtained under adiabatic elimination of the cavity amplitude. In this long-time regime the spin ensemble is dephased and it progressively saturates with time such that after some transient the system evolves to its ultimate stationary state that strongly depends on the value of the driving amplitude.

We then explored in detail the effect of a critical slowing-down of the cavity population near two phase transitions  represented by saddle-node bifurcations. This effect is characterized by a power law divergence of the transient time when the value of the driving amplitude approaches the critical one at which the phase transition occurs. We also calculated the smallest exponent of the asymptotic behavior of the system as it settles to the upper or lower stationary states. Their values turned out to be substantially smaller than the radiative dephasing rate of the individual spins in a noticeable interval of driving amplitudes exhibiting a minimum value that vanishes exactly at the dissipative phase transitions.

\begin{acknowledgments}
We would like to thank Andreas Angerer, Himadri Dhar, William Munro, Kae Nemoto and Stefan Putz for helpful discussions and the European Commission under Project No.~NHQWAVE (MSCA-RISE 691209) for support.  M.Z. acknowledges financial support by the Austrian Science Fund (FWF) through Project No. F49-P10 (SFB NextLite) and the Doctoral Program CoQuS (W1210). Some of the computational results presented here have been achieved using the Vienna Scientific Cluster (VSC). 
\end{acknowledgments}


\begin{thebibliography}{99}
\bibitem{expl_norm_form}The behavior of dissipative systems in the vicinity of bifurcations can often be characterized by the so-called normal form which doesn't dependent on a particular structure of the underlying system \cite{Strogatz00}.
\end{thebibliography}
\end{document}